\documentclass[pra,aps,twocolumn,showpacs,superscriptaddress,floatfix,amsmath,amssymb,nofootinbib]{revtex4}
\usepackage{amsfonts}
\usepackage{graphicx} 
\newcommand{\beq}{\begin{equation}}
\newcommand{\eeq}{\end{equation}}

\newcommand{\beqa}{\begin{eqnarray}}
\newcommand{\eeqa}{\end{eqnarray}}

\begin{document} 
\title{Quantum simulation of a triatomic chemical reaction with ultracold atoms on a waveguide}
\author{E. Torrontegui}
\affiliation{Departamento de Qu\'{\i}mica-F\'{\i}sica, Universidad del Pa\'{\i}s Vasco - Euskal Herriko Unibertsitatea, 
Apdo. 644, Bilbao, Spain}
\author{A. Ruschhaupt}
\affiliation{Institut f\"ur Theoretische Physik, Leibniz
Universit\"{a}t Hannover, Appelstra\ss e 2, 30167 Hannover,
Germany}
\author{D. Gu\'ery-Odelin}
\affiliation{Laboratoire Collisions Agr\'egats R\'eactivit\'e, CNRS UMR 5589, IRSAMC, Universit\'e Paul Sabatier, 118 Route de Narbonne, 31062 Toulouse CEDEX 4, France}
\author{J. G. Muga}
\affiliation{Departamento de Qu\'{\i}mica-F\'{\i}sica, Universidad del Pa\'{\i}s Vasco - Euskal Herriko Unibertsitatea, 
Apdo. 644, Bilbao, Spain}
\begin {abstract}
We study the scaling and coordinate transformation to physically simulate quantum three-body collinear chemical reactions of the type A$+$BC $\rightarrow$ AB+C by  the motion of single ultracold atoms or a weakly interacting Bose-Einstein condensate on an $L$-shaped waveguide. As an example we show that the parameters to model the reaction F+HH $\to$ H+HF with lithium are at reach with current technology. This mapping provides also an inverse scattering tool to find an
unknown potential, and 
a way to transfer the knowledge on molecular reaction dynamics
to design beam splitters for cold atoms with control of the channel outcome and vibrational excitation.      
\end{abstract}  	
\maketitle
{\it{Introduction}}. Ultracold atoms and ions are  
relatively easy to isolate, prepare, manipulate and detect  
by means of highly controllable operations that preserve their quantum coherence in the time scale of processes of interest. They have thus become natural candidates for performing physical, rather than numerical, simulations in which the effective Hamiltonians governing their dynamics can be made equal to the Hamiltonians of very different, simulated, quantum systems.
These simulations are thus based on a formal analogy and may   
predict the behavior of the simulated system under  
conditions hard to realize and/or calculate in the original one. 
The simulating system may also be interesting in its own,  
beyond the parameters relevant for the simulation, 
and lead to genuinely new phenomena and applications \cite{Dirac}.     
This opens exciting perspectives for many-body physics \cite{BDZ08}, and also for 
few-body systems.

In this letter we show that this approach can be applied to molecular dynamics and chemical reactivity by studying the analogy between reactive collinear three-body 
chemical reactions and the motion of a single cold atom, 
or possibly a weakly interacting condensate, on a potential surface
designed by a magnetic or optical waveguide. We put the emphasis on the chemical reaction, but the same procedure may shed light also on non-reactive collisions.   
What we propose and what is facilitated by ultracold atoms is basically a quantum dynamical version of the rolling ball analogy of chemical reactions with the ball ensemble substituted by 
a condensate or an ultracold-atom wavepacket, and the mechanical model potential by a 
magnetic or optical waveguide. 
Quantum effects are important for state to state 
(rather than averaged) results as well as for reactions involving a light atom transfer such as hydrogen. 

Most chemical reactions occur with steric requirements, i.e., a preferred direction of attack. 
The collinear configuration for the reaction path corresponds in many ``abstraction'' reactions involving halogen and alkali atoms to the lowest potential barrier and to the preferred orientation within a narrow cone of acceptance \cite{Levine}. 
Moreover collinear reactions may be induced by orienting cold  polar molecules with strong electric fields via the second order Stark effect \cite{Levine}.  
They are also a standard workbench for testing new calculational methods, examining the range of validity of several approximate theories, and exploring parameter variations over a wide range of values, difficult to implement with full 3D calculations. Accurate quantum calculations involve two mathematical coordinates and are still time consuming and specially troublesome when heavy atoms or high energies are involved. The results of interest are usually the branching ratios among the channels or the distribution of produced molecules among the possible vibrational states.

{\it Simulation Setting.}
The collinear chemical reaction
A+BC $\rightarrow$ AB+C, corresponds to the collision of an atom A and a nonrotating diatomic molecule BC with the three atoms aligned.  
We assume that the Born-Oppenheimer approximation holds 
and separate the fast electronic and the slow nuclear motions. 
In terms of nuclear masses, positions and momenta in a laboratory frame the nuclear motion is governed by the quantum-mechanical Hamiltonian  
\begin{eqnarray}
H=\underbrace{\frac{p_A^2}{2m_A}+\frac{p_B^2}{2m_B}+\frac{p_C^2}{2m_C}}_{\cal T}
+ V (x_A,x_B,x_C),
\label{haminit}
\end{eqnarray} 
where 
$V$ is the effective interaction between the three nuclei. 
The first step is the transformation from the ``chemical
reaction'' variables to the atomic ``simulation variables''. A second
important task is to show that the required parameters for the cold
atom experiment are available with current technology.

{\it{Mass-weighted coordinate system.}}
Let us introduce the center of mass (CM) coordinate
$R_{CM}:=(m_A x_A+m_B x_B+m_C x_C)/{M}$,
where $M:=m_A+m_B+m_C$, and the relative coordinates  
%
$
q_1:=x_B-x_A,\; q_2:=x_C-x_B.$
%
The corresponding momentum operators
are 
%
$P_{CM}:= \frac{\hbar}{i} \frac{\partial}{\partial R_{CM}}
= p_A + p_B + p_C$,
$P_{q_1} := \frac{\hbar}{i} \frac{\partial}{\partial q_1} 
= \frac{m_A}{M} P_{CM} - p_A$,
and 
$P_{q_2} := \frac{\hbar}{i} \frac{\partial}{\partial q_2} 
= p_C - \frac{m_C}{M} P_{CM}$.
%
The kinetic energy  ${\cal T}$ is not diagonal in terms of
them. To diagonalize ${\cal T}$, we use the mass-weighted coordinates \cite{Levine}
\beqa
\label{trans_position2}
Q_{1}&=&(aq_1+bq_2\cos\beta)/(\sqrt{\widetilde m} l),
\nonumber \\
Q_{2}&=&bq_2\sin\beta/(\sqrt{\widetilde m} l), 
\eeqa
with mass factors 
%
$a=\sqrt{{m_A(m_B+m_C)}/{M}}$, $b=\sqrt{{m_C(m_B+m_A)}/{M}}$,
$\tan\beta = \sqrt{{m_B M}/(m_A m_C)}$, 
%
and  scaling parameters $\widetilde m$ and $l$ that 
we can choose freely. 
The corresponding momentum operators are
\begin{eqnarray}
P_{Q_1} &:=& \frac{\hbar}{i} \frac{\partial}{\partial Q_1} 
= \frac{la\sqrt{\widetilde{m}}}{m_B+m_C}
\left(-\frac{m_B+m_C}{m_A} p_A 
+ p_B+p_C\right)\nonumber\\
P_{Q_2} &:=& \frac{\hbar}{i} \frac{\partial}{\partial Q_2} 
= b l \sin\beta \sqrt{\widetilde{m}}
\left(\frac{1}{m_C} p_C - \frac{1}{m_B} p_B\right),
\label{trans_momentum}
\end{eqnarray}
and the kinetic energy ${\cal T}$ takes the form 
\begin{eqnarray*}
{\cal T}=\frac{1}{2 M} P_{CM}^2 + \frac{1}{2\widetilde m l^2}
\left(P_{Q_1}^{2}+P_{Q_1}^{2}\right),
\end{eqnarray*}
%
The connection between
the simulation variables $\{R_{CM},Q_1,Q_2\}$ and the chemical reaction
variables $\{x_A,x_B,x_C\}$ is given by
%
\begin{eqnarray}
x_A&=&R_{CM} - \frac{l
  \sqrt{\widetilde{m}}}{m_A} Q_1,
\nonumber\\
x_B&=&R_{CM} + b l \sqrt{\widetilde{m}}
\left(\frac{\cos\beta}{m_C} Q_1 - \frac{\sin\beta}{m_B} Q_2\right),\nonumber\\
x_C&=&R_{CM} + \frac{b l \sqrt{\widetilde{m}}}{m_C} (\cos\beta Q_1 + \sin\beta
Q_2).
\label{trans_position}
\end{eqnarray}
In the following we ignore the trivial center of mass motion and
assume that the 
potential does only depend on the relative differences between the particle
positions. Then the corresponding time-dependent Schr\"odinger equation
associated with the Hamiltonian (\ref{haminit}) in the new variables is
\beq
\label{S2}
i\hbar\frac{\partial\Psi}{\partial \tau}
=-\frac{\hbar^2}{2\widetilde m}\left(\frac{\partial^2}{\partial Q_{1}^{2}}+\frac{\partial^2}{\partial Q_{2}^{2}}\right)
\Psi+ V_{Q} (Q_1,Q_2) \Psi,
\eeq
where we have set $\tau = t/l^2$ and
\begin{eqnarray}
V_Q (Q_1, Q_2) = l^2 V_q(q_1,q_2)=l^2 V(x_A,x_B,x_C).
\label{t4}
\end{eqnarray}
Equation (\ref{S2}) is the important result for the simulation, and describes  
2D quantum motion of a quantum particle of mass $\widetilde{m}$ on the potential   
$V_Q$. 


{\it{Potential energy surface.}}
We now specify the potential surface $V_q$ for the interaction between the three particles of the reaction. 
This might be an ab initio or, more generally, a semiempirical potential. 
Here we assume the semiempirical London-Eyring-Polanyi-Sato (LEPS) surface
\cite{sato, Muc, JW},  
\beqa
\label{potential}
\lefteqn{V_q(q_1,q_2) =}&&\\
& &\frac{1}{1+\Delta}\left[\sum_{i=1}^{3} U_i
- \sqrt{\sum_{i=1}^3 \alpha_{i}^{2}+ -\alpha_1\alpha_2-\alpha_2\alpha_3-\alpha_1\alpha_3}\right],\nonumber
\eeqa
where\\
%
$U_i=\frac{1}{4}D_i\left[(3+\Delta)e^{-2\beta_i(q_i-q_{i0})}
-(2+6\Delta)e^{-\beta_i(q_i-q_{i0})}\right]$,\\
$\alpha_i=\frac{1}{4}D_i\left[(1+3\Delta)e^{-2\beta_i(q_i-q_{i0})}
-(6+2\Delta)e^{-\beta_i(q_i-q_{i0})}\right]$,\\
and $q_3= q_1+q_2$. 
$D_i$, $\beta_i$ and $q_{i0}$ are the dissociation energy, the Morse parameter and the equilibrium
distance of the {\itshape i}-th  diatomic molecules that we can construct from the three atoms. The adjustable parameter $\Delta$ is optimized for 
each reaction. 
In the asymptotic regions, before and after the reaction happens, one of the atoms is far from the others and the potential energy is the one of a diatomic molecule  \cite{sato}. In the LEPS surface, this is given 
by the 
Morse function
\beq
V_j(q_j)=D_j[1-e^{-\beta_j(q_j-q_{j0})}]^{2},
\eeq
where  $j=1$ for the products' channel with the diatomic molecule AB,  or $j=2$ for the reactants' channel with the 
diatomic molecule BC. This potential near the equilibrium distance $q_{j0}$ can be harmonically approximated by
\beq
\label{harq}
V_j(q_j)=\frac{1}{2}K_j(q_j-q_{j0})^2,
\eeq
where $K_j=2D_j\beta_{j}^2$ is the force constant.  

Applying Eqs. (\ref{trans_position}) to the potential in the asymptotic regime
where $V_q (q_1, q_2) \approx V_j (q_j)$, we obtain for the
simulating frame that the energy surface in the asymptotic regions of the
products' and reactants' channels are, taking into account Eq. (\ref{t4}),
\beq
\label{harQ}
V_{Q}(Q_1,Q_2) 
\approx \frac{1}{2} \widetilde K_j\left[\chi_j(Q_1,Q_2)-\chi_{j,0}\right]^2, 
\eeq
where we have defined for the products' channel $(j=1)$
\begin{eqnarray*}
\chi_{1,0}&=& q_{10}\frac{a\sin\beta}{l \sqrt{\widetilde m}},\;
\widetilde K_1=\frac{K_1\widetilde ml^4}{a^2\sin^2\beta},\\
\chi_1(Q_1,Q_2)&=&\sin\beta\, Q_1 - \cos\beta\, Q_2 
=\frac{a \sin\beta}{l \sqrt{\widetilde m}}q_1,
\end{eqnarray*}
%
whereas for the reactants' channel $(j=2)$
\begin{eqnarray*}
\chi_{2,0}&=&q_{20}\frac{b\sin\beta}{l\sqrt{\widetilde m}},\;
\widetilde K_2=\frac{K_2\widetilde ml^4}{b^2\sin^2\beta},\\
\chi_2(Q_1,Q_2)&=&Q_2=\frac{b \sin\beta}{l \sqrt{\widetilde m}}q_2.
\end{eqnarray*}
The function $\chi_1$ is a rotation in the $(Q_1,Q_2)$ plane, so the
potential Eq. (\ref{harQ}) is, for the products, 
simply a rotated harmonic oscillator in the
$(Q_1,Q_2)$ plane.
In terms of the oscillation frequencies of the diatomic molecules $\nu_j$, 
the frequencies $\widetilde\nu_j$ of the harmonic
oscillators in Eq. (\ref{harQ}) are
\beq
\label{nu}
\widetilde\nu_1=\frac{l^2\sqrt{\mu_{AB}}}{a\sin\beta}\nu_1,
\quad 
\widetilde\nu_2=\frac{l^2\sqrt{\mu_{BC}}}{b\sin\beta}\nu_2,
\eeq
where $\mu_{AB}$ and $\mu_{BC}$ are reduced masses for the diatomic molecules.  
The value of $l$ can be fixed from these last equations, so that  
the potential parameters of the simulation can be made realistic.


{\it{Initial atomic velocity.}}
To set the initial velocity
of the cold atom $v_{Q_1}$ in the reactants' channel we first estimate the velocities
involved in the chemical reaction. 
If the reaction
happens at temperature $T$ the rms mean velocities 
for the atom A and diatomic molecule BC 
along a given direction are respectively $(k_BT/m_A)^{1/2}$ and $[k_BT/(m_B+m_C)]^{1/2}$, where $k_B$ is
the Boltzmann constant. We may then assume   
\beq
v_C \approx v_B, \; v_A-v_B=\sqrt{k_BT}\bigg(\frac{1}{\sqrt{m_A}}+\frac{1}{\sqrt{m_B+m_C}}\bigg), 
\eeq
which, following from Eqs. (\ref{trans_momentum}),
corresponds to the atom velocity   
\begin{eqnarray}
\label{Q1}
v_{Q_1}&=& \frac{l a}{\sqrt{\widetilde m}} (v_B-v_A)\nonumber\\
&=& a l \sqrt{\frac{k_{B}T}{\widetilde
    mm_A}}\bigg(1+\sqrt{\frac{m_A}{m_B+m_C}}\bigg),\\
v_{Q_2} &\approx& 0.
\end{eqnarray}


{\it{Example and numerical values.}}
As an explicit example 
we consider the reaction F+H$_2\rightarrow$ FH+H, where F$\rightarrow$A, H$\rightarrow$B,
and H$\rightarrow$C, so $m_A=3.15\cdot 10^{-26}$ kg and $m_B=m_C=1.66\cdot 10^{-27}$ kg. 
For this particular reaction, $\Delta=0.164$ \cite{Muc}, 
and the mass factors are 
$a=5.48\cdot 10^{-14}$, $b=3.98\cdot 10^{-14}$ and $\beta=46.45^{\circ}$. 
For the diatomic molecule HF, $q_{10}=0.917 \AA$, 
the dissociation energy is $D_1=9.609\cdot 10^{-19}$ J and the Morse parameter $\beta_1= 2.242\AA^{-1}$, so the force constant is $K_1=2D_1\beta_{1}^{2}= 966 $ N/m. Consequently the oscillation frequency
$\nu_1=(K_1/\mu_{HF})^{1/2}/2\pi= 1.246\cdot 10^{14}$ Hz
whereas for HH,  $q_{20}=0.742$, $D_2= 7.608\cdot 10^{-19}$ J,
$\beta_2= 1.942\AA^{-1}$ and $K_2= 573.85 $ N/m so
$\nu_2=(K_2/\mu_{HH})^{1/2}/2\pi= 1.32\cdot 10^{14}$ Hz. 

To simulate the reaction we propose 
$^{7}$Li atoms. One advantage of $^{7}$Li is that the interatomic repulsive interactions are extremely tunable with a Feshbach resonance. The zero crossing of the $s$-wave scattering length is the shallowest known, so that only modest field stability is needed to achieve a non-interacting gas \cite{Litio1}.
We thus have $\widetilde m=1.1526\cdot 10^{-26}$ kg and 
set  
$l=6.55\cdot 10^{-6}$. Defining the valley depths $\widetilde V_j=D_{j}l^2$ ($j=1, 2$) 
according to Eq. (\ref{t4}), the parameters in the asymptotic region for the reactants' channel are 
%
$
\widetilde\nu_2=5.66\, {\rm kHz}$, and $\widetilde V_2=2.4 \,\mu {\rm K}$, 
%
whereas in the asymptotic region of the products' channel, 
%
$
\widetilde\nu_1=5.34\, {\rm kHz}$, and $\widetilde V_1=3\,\mu {\rm K}$, 
%
The choice of a light atom such as lithium is also dictated by the requirement of achievable transverse frequencies in the reactants' and products' channels with standard techniques (see below). To illustrate the scaling of distances and velocities note 
that a displacement
of $1\AA $ of the atom F along the reactants' channel corresponds to a displacement
of $7.8$ $\mu$m of the lithium atom according to Eqs. (\ref{trans_position2}). If the reaction 
occurs at room temperature, $T=298$ K, Eq. (\ref{Q1}) sets 
for the lithium atom a velocity $v_{Q_1}=5$ mm/s along the asymptotic region of the reactants' channel.
%
\begin{figure}[t]
\begin{center}
\includegraphics[height=5cm ,angle=0]{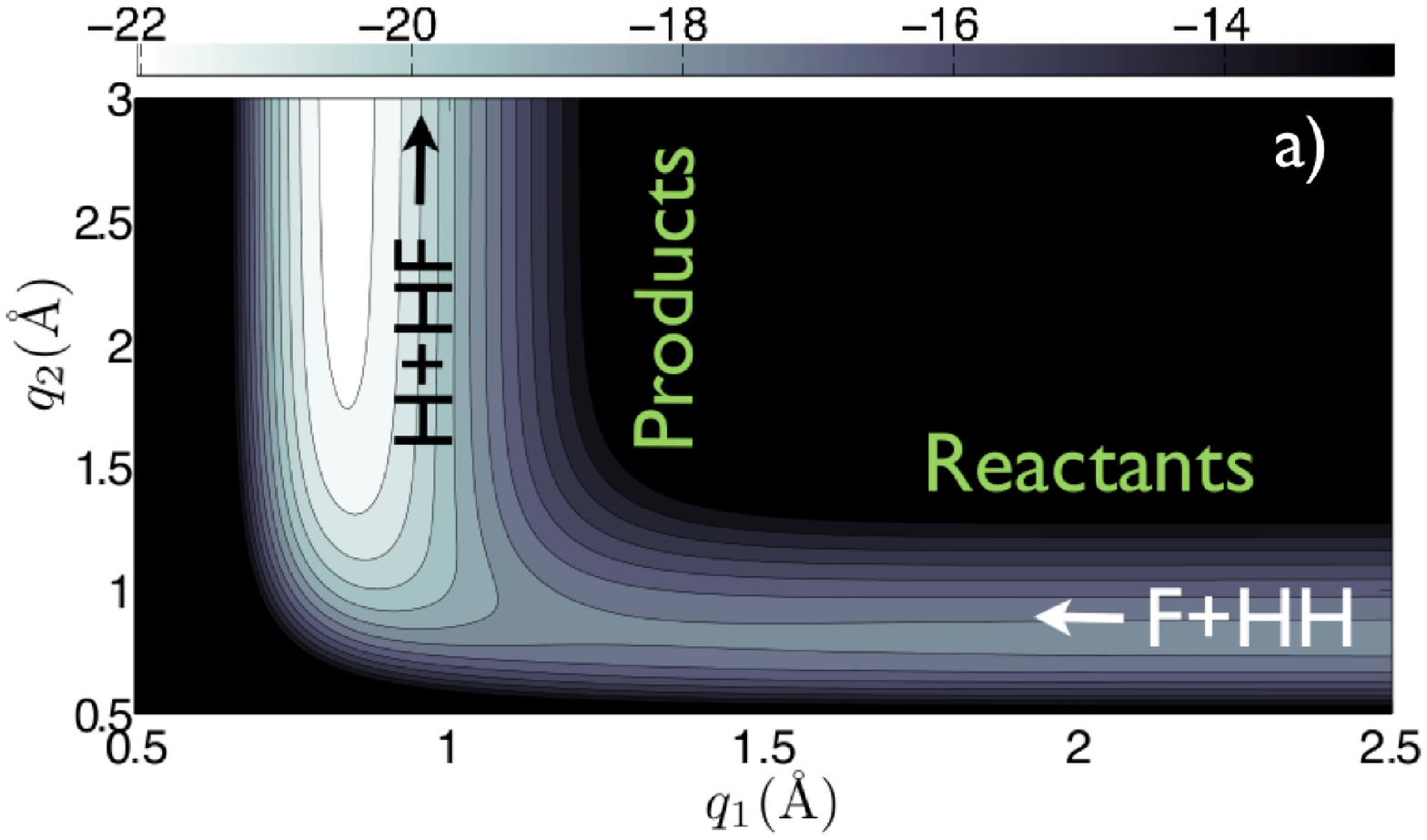}\vspace{0.01cm}
\includegraphics[height=4.7cm,angle=0]{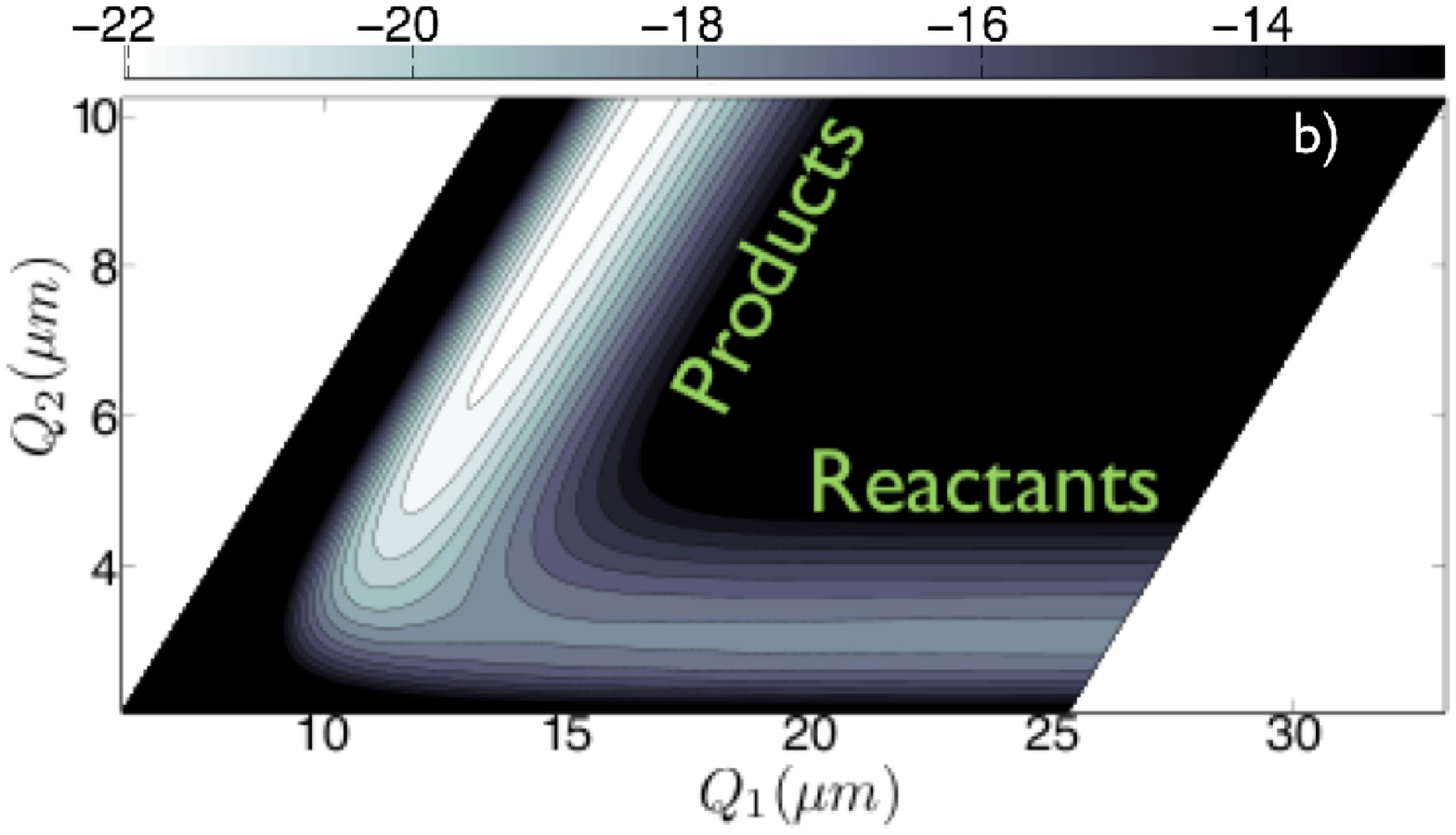}
\end{center}
\caption{\label{pot}
(Color online) (a) Contour map of the potential energy surface  (\ref{potential}) for $H_2+F\rightarrow HF+H$. 
(b) Contour map of the potential for the $^{7}$Li atom that simulates 
the chemical reaction. 
In both cases the energy is in units of the zero point energy of the reactants' valley, and the surface is truncated well below zero energy (the asymptotic value when all atoms are far apart) to better visualize the saddle and reaction path.   
}
\end{figure}
%
The control of matter waves at such low velocities is at reach \cite{Litio2}.   
In Fig. \ref{pot} we plot the potential energy of the chemical reaction H$_2$+F $\rightarrow$ HF+H given 
by Eq. (\ref{potential}) to see the transformations from the chemical reaction
parameters $\{q_1, q_2\}$ into
the ``laboratory'' simulation waveguide on which the $^{7}$Li atom moves. 
Note the advanced saddle, and the deeper product's valley, responsible for the
exoergicity and the vibrational excitation of the resulting HF molecule.    

The experimental realization with ultracold atoms involves (I) the preparation of a propagating matter wave in a guide and (II) the realization of a guide with the appropriate shape.  A Bose-Einstein condensate, rather than the repetition of the experiment with single atoms, provides the ideal setting since the fate of the whole quantum wave packet can be measured in one single experiment. The propagation of a Bose-Einstein condensate into straight magnetic or optical guides has already been demonstrated experimentally \cite{BBD01,LCK02}.  More recently, the production of guided atom lasers shows that a large control of the matter wave parameters such as the mean velocity (5-30 mm/s), the transverse mode occupations, the internal state, or the linear atomic density can be achieved \cite{GRG06,CJK08,GCJ09,KWE10}. Using different outcoupling mechanisms, the matter wave can be prepared in the transverse ground state \cite{CJK08,GCJ09}. In these latter schemes, the diluteness of the matter wave suppresses the role of interactions providing a well-suited system for the quantum scattering experiments of interest, without the need of Feshbach resonance tuning. The second aspect deals with the potential modeling to design simple reactive chemical reactions. Different strategies can be envisioned (i) with wires sculptured on atom chips by a focused atom beam technique \cite{Rei02,DAV07}, (ii) with adiabatic radio-frequency potentials \cite{MCL06,LSH06}, (iii) with high resolution time averaged optical potentials ``painted'' by a tightly focused rapidly moving laser beam on a 2D canvas formed by a static light sheet \cite{HRM09}. A canvas of 60 $\mu$m diameter, and a radial condensate thickness of less than 1 $\mu$m as the ones realized in \cite{HRM09}, are enough for the spatial range and resolution needed for the 
simulation, see Fig. \ref{pot}b. Moreover   
the potential depth can be controlled by velocity or intensity modulation, and no decrease in the number of condensate atoms is observed after 2.5 s, again more than enough for 
implementing a process of the order of milliseconds. Reaction probabilities could 
be detected with an \emph{in situ} high resolution imaging, whereas the coherent vibrational excitation is measurable after a few ms time-of-flight.     
A high flexibility in the guide design is also provided by combining properly these various techniques and/or using time-dependent optical or magnetic potentials \cite{MCW00,KFA01}. The simplest realization would involve a crossed red-detuned dipole beams configuration in combination with a well positioned repulsive potential wall realized by a sheet of blue-detuned laser light \cite{diode}. Alternatively, one could study the motion of an ion into a well-designed guide. Ultracold ions have already been transported in complex structures \cite{BOV09,SPM10}, 
but their propagation in guides has not been investigated so far.


{\it{Discussion and Outlook.}}
We have worked out the mapping between a quantum-mechanical collinear triatomic chemical reaction and the motion of ultracold atoms  on a tilted, $L$-shaped waveguide. As an example we have shown that  
the parameters for simulating the reaction F $+$ H$_2$ $\to$ FH $+$ H using $^{7}$Li 
can be implemented with currently available technology. This approach is thus complementary to other proposals for simulating chemical reactions \cite{kassal.2008}, which are more ab initio and do not need any previous calculation of the potential surface or application of the Born-Oppenheimer approximation,  
but require a quantum computation with hundreds of coherently manipulated qubits. This is currently out of reach for a reaction like the one discussed. The present approach is less fundamental, since it assumes a potential surface and the Born-Oppenheimer approximation to hold, but also easier to implement.
As an inverse scattering tool,   
the capability to manipulate the potential parameters may be used to fit
experimental results of the chemical reaction and find the right potential.

By a straightforward generalization, we could also simulate collinear
four-atom reactions by an ultra-cold atom in a three-dimensional potential.
As a further application of the mapping, the vast knowledge and experience accumulated on chemical reaction dynamics,
in particular for triatomic systems in the collinear configuration, is now ready to be transferred to design 
crossed laser beams or waveguide bends with different properties. They could be used, for example, as control devices for asymmetrical beam splitting into the channels or for controlling the transverse vibrational excitation. An example of this is the 
recent design of an atom diode or one-way barrier \cite{diode}.

We acknowledge the kind hospitality of the Max Planck Institute
for the Physics of Complex Systems in Dresden, funding
by the Basque Government (Project No. IT 472-10), Ministerio de
Ciencia e Innovaci\'on (Project No. FIS2009-12773-C02-01), R\'{e}gion Midi-Pyr\'{e}n\'{e}es, Institut Universitaire de France and Agence Nationale de la Recherche (Project No. ANR-09-
BLAN-0134-01). E.T. acknowledges support from the Basque
Government (Grant No. BFI08.151).

\end{document}